\documentclass{aa}  

\usepackage{graphicx}
\usepackage{natbib}
\usepackage{subfig}
\usepackage{epsfig} 
\usepackage{ifpdf} 
\usepackage{fancyvrb}
\usepackage{txfonts}
\usepackage{xcolor}
\usepackage{soul}
\usepackage{mathtools}

\begin{document} 

\title{The relentless variability of Mrk\,421 from the TeV to the radio}

\author{ 
A. Arbet-Engels\inst{1} \and D. Baack\inst{2} \and M. Balbo\inst{3} \and 
A. Biland\inst{1} \and M. Blank\inst{4} \and T. Bretz\inst{1,5} \and 
K. Bruegge\inst{2} \and M.\,Bulinski\inst{2} \and J. Buss\inst{2} \and M. Doerr\inst{4} \and 
D. Dorner\inst{4} \and D. Elsaesser\inst{2} \and D. Hildebrand\inst{1} \and 
K. Mannheim\inst{4} \and S.\,A.\,Mueller\inst{1} \and D.\,Neise\inst{1} \and M. Noethe\inst{2} \and A. Paravac\inst{4} \and W. Rhode\inst{2} \and 
B. Schleicher\inst{4} \and 
K. Sedlaczek\inst{2} \and 
A. Shukla\inst{4} \and
V. Sliusar\inst{3}\thanks{Corresponding authors, e-mails:  vitalii.sliusar@unige.ch \& roland.walter@unige.ch} \and R. Walter\inst{3}\footnotemark[1] \and 
E.\,von\,Willert\inst{4}
}

\institute{
Department of Physics, ETH Zurich, Otto Stern Weg 5, CH-8093 Zurich, Switzerland \and 
Fakult\"at Physik, Technische Universit\"at Dortmund, Otto Hahn Str. 4a, D-44227 Dortmund, Germany \and 
Department of Astronomy, University of Geneva, Chemin d'Ecogia 16, CH-1290 Versoix, Switzerland \and 
Institut f\"ur Theoretische Physik und Astrophysik, Universit\"at W\"urzburg, Emil Fischer Str. 31, 97074 W\"urzburg, Germany \and 
Physikalisches Institut III A, RWTH Aachen University, Otto Blumenthal Str., D-52074 Aachen, Germany
}

\date{Received Month Day, Year; accepted Month Day, Year.}

\abstract
   {The origin of the $\gamma$-ray emission of the blazar \object{Mrk\,421} is still a matter of debate.}
   { We used 5.5 years of unbiased observing campaign data, obtained using the FACT telescope and the Fermi LAT detector  at TeV and GeV energies, the longest and densest so far, together with contemporaneous multi-wavelength observations, to characterise the variability
of \object{Mrk\,421} and to constrain the underlying physical mechanisms. }
   {We studied and correlated light curves obtained by ten different instruments and found two significant results.}
   {The TeV and X-ray light curves are very well correlated with a lag of $<0.6$ days. The GeV and radio (15 Ghz band) light curves are widely and strongly correlated. Variations of the GeV light curve lead those in the radio.
   }
   {Lepto-hadronic and purely hadronic models in the frame of shock acceleration predict proton acceleration or cooling timescales that are ruled out by the short variability timescales and delays observed in Mrk\,421. Instead the observations match the predictions of leptonic models. }

\maketitle

\section{Introduction\label{sec:introduction}}

Blazars are active galactic nuclei (AGNs) emitting a relativistic jet aligned with the observer's line of sight. Their spectra consist of two broad non-thermal components peaking in wavelengths from radio to optical, and in the $\gamma$-rays, respectively. If the low-energy component is very likely emitted by electron synchrotron, the origin of the high-energy component is still a matter of debate. 

In some objects, including Mrk\,421, the $\gamma$-ray emission is so rapidly variable that moderate Doppler boosting is inescapable during fast flares \citep{1998MNRAS.293..239C}. The fastest variations suggest that either a shock moves relativistically within the jet or that the emission region is much smaller than the gravitational radius and driven by interactions between stars/clouds and the jet, or by magnetic reconnection \citep{2017ApJ...841...61A}.

Mrk\,421 is a high-synchrotron-peaked blazar ($z = 0.031$), with a low-energy synchrotron component peaking above $10^{17}$\,Hz, featuring bright and persistent GeV and TeV emission with frequent flaring activities. Its average spectral energy distribution has been modelled with a one-zone leptonic synchrotron self-Compton model \citep[SSC;][]{abdo_2011ApJ...736..131A} or with a hadronic model where the accelerated protons cool through synchrotron emission \citep{2015MNRAS.448..910C} or interact with the leptonic synchrotron photons to create a cascade of pions and muons, decaying into $\gamma$-rays and neutrinos \citep{2001APh....15..121M,2001ICRC....3.1153M}.

Numerous multi-wavelength campaigns have been carried out \citep[e.g.][]{2004ApJ...601..759T,2015A&A...576A.126A}. During the 2009 campaign, Mrk\,421 was simultaneously observed from the radio to the TeV band for 4.5 months in the absence of strong flares \citep{2015A&A...576A.126A}. The fractional variability revealed that, despite Mrk\,421 being in a relatively mild state, most variability lies in the X-ray ($F_{var}=0.5$) and TeV ($F_{var}=0.3$) bands. A harder-when-brighter behaviour was found in the X-rays. The smallest variations were observed in the radio. \cite{2015A&A...576A.126A} determined a positive cross-correlation between the fluxes at keV and TeVs at zero time lag, with a maximum lag of about five days, determined from the z-transformed discrete correlation functions. Cross-correlations between optical/UV and X-rays did not show any significant correlation. While the correlation studies between TeV and keV fluxes have typically been consistent with a zero time lag, \cite{2008ApJ...677..906F} reported a time lag of $2.1 \pm 0.7$ ks between these two bands during strong flaring activity of Mrk\,421 in March 2001.

In this paper, we are using a 5.5-year  observational campaign conducted on Mrk\,421 at TeV energies with the First G-APD Cherenkov Telescope \citep[FACT;][]{2013JInst...8P6008A}. These observations were not triggered (the observations were regularly scheduled independently of the source activity) and are therefore unbiased, and as regular as possible taking into account observing conditions and technical constraints. 
We also use continuous radio, optical, ultraviolet, X-ray, hard X-ray, and GeV light curves obtained quasi-simultaneously to the FACT campaign by eight additional instruments with the aim being to improve the observational constraints and compare them with the predictions of various emission models. 

Section \ref{sec:data} describes the various instruments and data used. The multi-wavelength variability of Mrk\,421 is studied in Sect. \ref{sec:timing} and compared to model expectations in Sect. \ref{sec:discussion}. Our conclusions are summarised in Sect. \ref{sec:conclusions}.

\section{Data and analysis\label{sec:data}}

\begin{figure}
\centering
\includegraphics[width=\columnwidth]{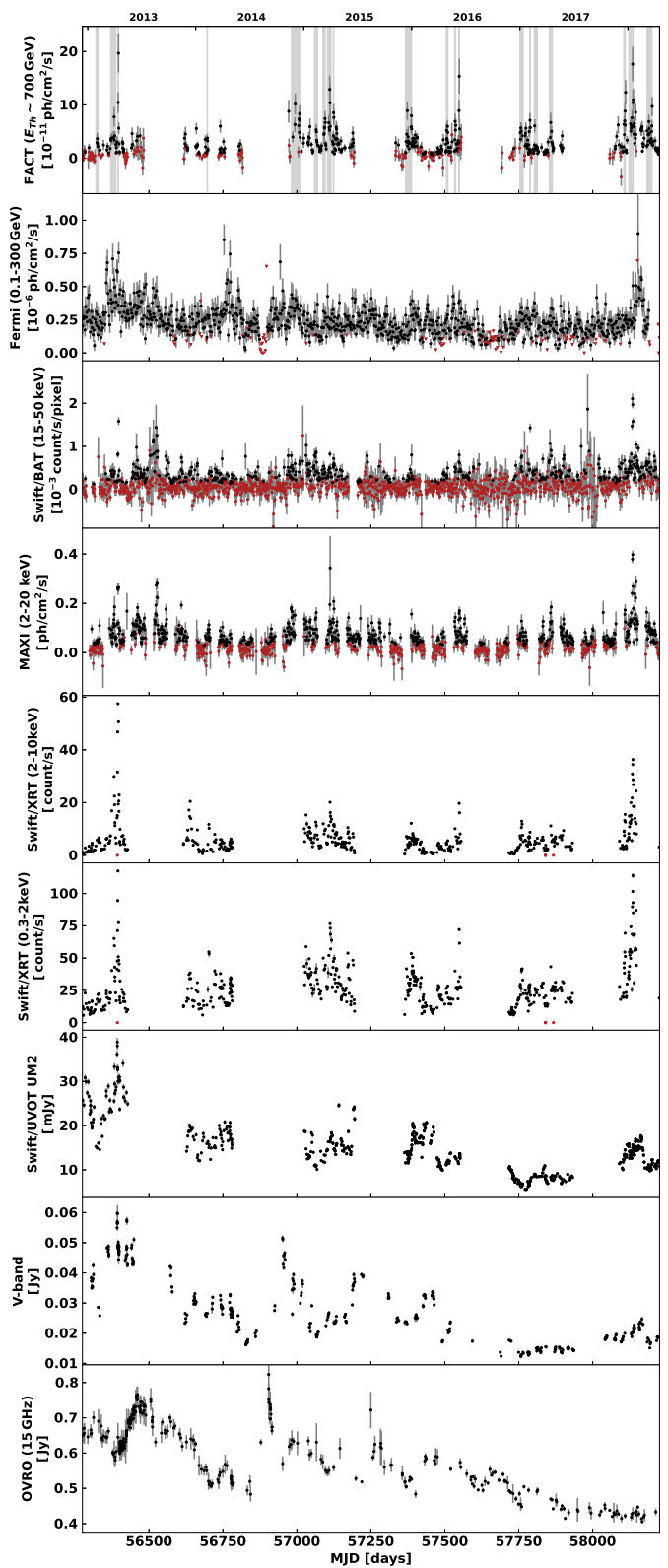}
\caption{Long-term light curves of Mrk\,421 obtained from the radio to the TeV between December 14, 2012, and April 18, 2018. From top to bottom: FACT ($E_{Th} \sim 700$\,GeV), Fermi LAT (0.1-300\,GeV), MAXI (2-20\,keV), Swift/BAT (15-50\,keV), Swift/XRT (2-10\,keV), Swift/XRT (0.3-2\,keV), Swift/UVOT UVM2, V-band optical observations, and radio observations at 15\,GHz. Flux measurement errors are denoted as vertical grey bars. Points with signal-to-noise ratio lower than two are coloured red (see Sect.~\ref{sec:data} for details on the cleaning procedures). For the FACT light curve, there are 177 (out of 580) such points. For the Fermi LAT light curve, 95\% flux upper limits (TS<25) are shown with triangles. For the FACT light curve, the flares listed in Table~\ref{tab:flares} are highlighted in grey.}
\label{fig:LC}
\end{figure}

To characterise the broadband variability of Mrk\,421,
we used data from nine different instruments spanning from the radio to the TeV and obtained between December 14, 2012, and April 18, 2018. The data and their analysis are presented in the following sections. 
The resulting light curves are presented in Fig.~\ref{fig:LC}. 
The light curves obtained for FACT, Monitor of All-sky X-ray Image (MAXI), and Swift Burst Alert Telescope (Swift/BAT) include negative fluxes, as expected when the source is not detected. The Fermi Large Area Telescope (LAT) light curve is always positive because of the use of a positively defined model in the maximum-likelihood fitting. Some of the analyses presented in Sect. \ref{sec:timing} disregard negative or low signal-to-noise-ratio ($<2 \sigma$) data points.
Mrk\,421 was observed at various flux states in all bands, with apparent flare duration roughly decreasing with energy \citep{sinha_2015A&A...580A.100S, hovatta_2015MNRAS.448.3121H, kapanadze_2017ApJ...848..103K}. 
Flares observed from the X-ray to the TeV are narrow enough to be identified individually (see Table~\ref{tab:flares}). At longer wavelengths the flares become wider and overlap. 

\subsection{FACT\label{sec:fact}} 
FACT is an imaging atmospheric
Cherenkov telescope with a 9.5\,m$^2$ segmented mirror located at the
Observatorio del Roque de los Muchachos on the island La Palma at an
elevation of 2.2\,km \citep{2013JInst...8P6008A}. FACT has been in operation
since October 2011 and is designed to detect gamma rays with energies
from hundreds of GeV up to about 10\,TeV by observing the Cherenkov
light produced in extensive air showers induced by gamma and cosmic
rays in the Earth's atmosphere. The FACT telescope performs
observations in fully remote and automatic mode. At the beginning of a shift,
the telescope is started by a remote operator, usually from home. The telescope software 
takes over full control and monitoring of the telescope, executing the predefined schedule 
during the whole night of observations. In case of any problems or unfavourable weather conditions an operator is called to take appropriate action.
Such an approach allows long-term regular observations of bright TeV sources at low cost.
The excellent and stable performance of the camera uses silicon-based photosensors (SiPM, also known as Geiger-mode Avalanche Photo Diodes (G-APDs)) and a feedback system to keep the gain of the 
photosensors stable \citep{Biland_2014}, which helps to maximise the duty cycle of the instrument
and minimise the gaps in the light curves \citep{2017ICRC...35..609D}, allowing for 
observation with bright ambient light. Regular operations take place with light conditions up to about 20 times brighter than the dark night conditions. Observations are typically interrupted for 4-5 days every month. The telescope can operate during full-Moon conditions \citep{2013ICRC...33.1132K}.

The light curve of Mrk\,421 was obtained using the Modular Analysis and
Reconstruction Software
\citep[MARS, revision 19203\footnote{https://trac.fact-project.org/browser/trunk/Mars/};][]{2010apsp.conf..681B}. Thanks to the stable and homogeneous gain,
no additional calibration is needed. After signal extraction, the
images are cleaned using a two-step process. First, islands with an
arrival time difference of neighbouring pixels of less than 2.4\,ns
are formed and adjacent islands are joined. In a second step, islands
with a total signal of less than 25\,pe (photon equivalents) are rejected. A principal component analysis of the remaining distribution results in a set of parameters
describing the shower image. A detailed description of the data selection, reconstruction, and applied cuts for light curve and spectral extraction are provided in \cite{2019ICRC...358..630D}.  The signal from the source is determined
cutting at $\theta^2 < 0.037$\,deg$^2$, where
$\theta$ is the angular distance between the reconstructed source position and the real source position in the camera plane.
The source
is observed in wobble-mode \citep{1994APh.....2..137F} with a distance
of 0.6\,deg from the camera centre.
To calculate the
background, the same analysis is performed for five off-source regions, each located 0.6\,deg from the camera centre. The source excess rate is derived
by subtracting the scaled signal 
in these off-regions from the signal at the source position and dividing it by the effective exposure time.

The Crab Nebula is a standard candle at TeV energies. Even though flares 
have been seen at MeV/GeV energies, similar flux changes have not  
been found at TeV energies. Therefore, the excess rate of the Crab Nebula 
is used to study and correct the dependence of the excess rate from the 
zenith distance and trigger threshold \citep[changed with the ambient light 
conditions;][]{2013arXiv1308.1516B}. These dependencies are similar to those of the cosmic-ray 
rate described in \cite{2019APh...111...72B} and details can be found in 
\cite{2019ICRC...358..630D}.
For the studied data sample, the corrections in zenith distance are less 
than 10\% for more than 92\% of the nights and in trigger threshold less 
than 10\% for more than 72\% of the nights. The maximum correction in 
zenith distance is 47\% with only one night with a correction larger than 
45\%. In threshold, the largest correction is 69\%, but 98\% of the nights 
have a correction smaller than 60\%.
To take into account the influence of the different spectral slope of 
Mrk\,421 compared to the Crab Nebula, the spectra of 75 time ranges 
between January 2013 and April 2018 determined with the Bayesian Block 
algorithm (see Sect.~\ref{sec:timing}) were extracted (using the method described in \cite{2015ICRC...34..707T} and the cuts from \cite{2019ICRC...358..630D}) 
and fitted with a simple power 
law. Within the uncertainties, no obvious dependency was found between index and flux,
and so the harder-when-brighter behaviour reported in \cite{2015A&A...576A.126A} cannot be confirmed or rejected. The distribution of indices 
yields an average spectral slope of $3.35\pm0.23$, which is compatible with some
previously published results of other telescopes taking into account the 
different energy ranges and instrument systematic errors  \citep{2007ApJ...663..125A}.
Assuming different slopes from $3.12$ to $3.58$, the corresponding energy thresholds and 
integral fluxes were determined in order to estimate the systematic error of 
a varying slope of the spectrum of Mrk 421. This results in a systematic 
flux uncertainty of less than 12\% which was added quadratically to the 
statistical uncertainties.

For the light curve of Mrk\,421, a data sample from December 14, 2012,
until April 18, 2018 (MJD 56275 to 58226), has been selected applying a
data-quality selection cut based on the cosmic-ray rate
\citep{2017ICRC...35..779H}. For this, the artificial trigger rate above a threshold of 750\,DAC-counts, $R750,$ is calculated and selected to be independent of the trigger threshold. As described
in \cite{2017ICRC...35..612M} and \cite{2019APh...111...72B}, the
dependence of $R750$ on the zenith distance is determined and a
corrected rate, $R750_{\rm cor }$ , is calculated. To account for seasonal
changes of the cosmic-ray rate due to changes in the Earth's
atmosphere, a reference value, $R750_{\rm ref}$ , is determined for each moon
period. In a distribution of the ratio $R750_{\rm cor}$/$R750_{\rm ref}$,
the good-quality data can be described with a Gaussian distribution.
Data obtained during bad weather deviate from this. A cut is applied at the
points where the distribution of the data starts deviating from the
Gaussian distribution. Data with good quality are selected using a cut of $0.93 <
R750_{\rm cor}/R750_{\rm ref} < 1.3$. This results in a total Mrk\,421 data sample of 1628 hours of observational
data of good quality. This sample contains data from
649 nights with up to 7.5\,hours of observation per night. The average observation time is 2.5 hours per night. For the light curve, nights with an observation time of less
than 20\,minutes were rejected.
The FACT light curve of Mrk\,421 is presented in Fig.~\ref{fig:LC} (uppermost panel).

Systematic uncertainties can be checked by verifying the stability of the Crab Nebula, which was observed daily contemporaneously to Mrk\,421. When filtered for bad weather and high night-sky background periods, the differences between the measured Crab fluxes and the reference Crab flux, when normalised by the observational uncertainties, are well represented by a Gaussian distribution.

\subsection{Fermi LAT\label{par:fermi}}
The LAT onboard the Fermi Gamma-ray Space Telescope is the most sensitive $\gamma$-ray telescope in orbit to date. Since August 4, 2008, it has been observing the sky from 20 MeV to 300\,GeV~\citep{2009ApJ...697.1071A} using a charged particle tracker and a calorimeter (which reach a total of 8.7 radiation length), surrounded by a segmented anti-coincidence system. Its point spread function (PSF) strongly depends on energy, reaching a $68\%$ containment radius of $\sim 0.1^{\circ}$ at 40\,GeV \citep{2009APh....32..193A}. More information about the LAT is provided in \cite{2012ApJS..203....4A} and \cite {2012APh....35..346A}. 

For the purpose of this paper we analysed Fermi LAT data from August 4, 2008, to April 18, 2018. For the variability and correlation studies we only used the data from December 14, 2012, corresponding to the FACT observations. Data were reprocessed with the PASS8\footnote{http://fermi.gsfc.nasa.gov/ssc/data/analysis/documentation/\\/Pass8$\_$usage.html} pipeline and analysed using the Fermi Science Tool v10r0p5 package\footnote{http://fermi.gsfc.nasa.gov/ssc/data/analysis/software/}. In order to improve the quality of the data analysis, we selected only those photons belonging to \texttt{evclass=128} and \texttt{evtype=3}, flagged with \texttt{(DATA\_QUAL==1) \&\& (LAT\_CONFIG==1)}, taken while the spacecraft was outside the South Atlantic Anomaly and the source was in the field of view of the satellite. We used the \texttt{zmax=90}
option in the \texttt{gtltcube} tool to minimise the background due to the atmospheric $\gamma$-rays originating from the Earth's limb.

Given the energy-dependent PSF of the instrument \citep{2012ApJS..203....4A}, we considered only events with energy 100\,MeV $<$ E $<$ 300\,GeV, in a region of intereset (ROI) of $20^\circ$ centred on Mrk\,421. The fitting model, instead, included sources from the LAT four-year Point Source Catalog\footnote{http://fermi.gsfc.nasa.gov/ssc/data/access/lat/4yr\_catalog/} within $25^\circ$ of Mrk\,421. We left the normalisation and the gamma index parameters free to vary for those sources less than $10^\circ$ from the ROI centre and with a test statistic (TS) higher than 25 (corresponding to a $\sim 5 \sigma$ detection) or presenting a variability index higher than $72.44$, which suggests a monthly timescale variability of the source \citep{2015ApJS..218...23A}. We used \texttt{gll\_iem\_v06.fits}\footnote{http://fermi.gsfc.nasa.gov/ssc/data/access/lat/BackgroundModels} \citep{2016ApJS..223...26A} and \texttt{iso\_P8R2\_SOURCE\_V6\_v06.txt$^{5}$} to describe the diffuse Galactic $\gamma$-ray emission and the isotropic background component, respectively.

The best fit on the entire data sample, with a power-law model described as $dN/dE = N(E/E_{\rm scale})^{-\Gamma}$, returned a normalisation of $N=(2.007 \pm 0.012) \times 10^{-11}$ ph~cm$^{-2}$~s$^{-1}$~MeV$^{-1}$ and a gamma index $\Gamma=1.780 \pm 0.004$, using $E_{\rm scale}=1187$\,MeV. This implies a total integrated flux of $F_{0.1-300~\mathrm{GeV}}=(2.10 \pm 0.02) \times 10^{-7}~{\rm ph~cm^{-2}~s^{-1}}$. When Mrk\,421 was detected with TS<25, we calculated 95\% flux upper limits. Throughout this paper, we use a binning of two days, leading to upper limits for 9\% of the data points (a one-day binning would lead to upper limits for 33\% of the data points).  Figure~\ref{fig:LC} shows the Fermi light curve. The spectral indices derived for each time bin (average $1.79 \pm 0.16$) do not suggest any significant spectral variability during these observations.

Using higher TS criteria to fix the parameters of the other sources yielded a difference of $\lesssim 1.8\%$ on Mrk\,421 integral fluxes, and $\lesssim 0.5\%$ on the spectral index. Leaving the Galactic and isotropic normalisations free to vary or fixing them to the average ten-year value yields a final difference of $\lesssim 0.5\%$ on the integral fluxes and $\lesssim 0.03\%$ on the slope. 

\subsection{Swift/BAT\label{par:bat}}

The wide field of view of the Swift/BAT \citep{0067-0049-209-1-14} onboard the Swift satellite allows  the complete sky at hard X-rays to be monitored every few hours. For sources as bright as Mrk\,421, light curves can be obtained in different energy bands with a resolution of days.

The Swift/BAT reduction pipeline is described in \cite{2010ApJS..186..378T} and \cite{2013ApJS..207...19B}. Our pipeline is based on the BAT analysis software \texttt{HEASOFT} version 6.13. A first analysis was performed to derive background detector images. We created sky images (task \texttt{batsurvey}) in the eight standard energy bands (in keV: 14 - 20, 20 - 24, 24 - 35, 35 - 50, 50 - 75, 75 - 100, 100 - 150, 150 - 195) using an input catalogue of 86 bright sources that have the potential to be detected in single pointings. The detector images were then cleaned by removing the contribution of all detected sources (task \texttt{batclean}) and averaged to obtain one background image per day. The variability of the background detector images was then smoothed pixel-by-pixel fitting the daily background values with different functions (spline, polynomial). A polynomial model with an order equal to the number of months in the data set adequately represents the background variations.

The BAT image analysis was then run again using these smoothed averaged background maps. The new sky images were then stored in an all-sky pixel database by properly projecting the data onto a fixed grid of sky pixels, preserving fluxes (the angular size of the BAT pixels varies in the field of view). This database can then be used to build local images and spectra or light curves for any sky position. We compared the result of our processing to the standard results presented by the Swift team (light curves and spectra of bright sources from the Swift/BAT 70-months survey catalogue\footnote{http://swift.gsfc.nasa.gov/results/bs70mon/}) and  found very good agreement.

The Swift/BAT light curves of Mrk\,421 were built in several energy bands. For each time bin and energy band a weighted mosaic of the selected data is first produced and the source flux is extracted assuming fixed source position and shape of the PSF. The signal-to-noise ratio of the source varies regularly because of intrinsic variability, its position in the BAT field of view, and distance to the Sun. The 15-50\,keV one-day-bin light curve presented in Fig.~\ref{fig:LC} spans from December 14,  2012, to April  18,  2018, which is over 29344 orbital periods or almost 5.5 years.

\subsection{Swift/XRT\label{par:xrt}}

The Swift X-Ray Telescope \citep[XRT;][]{2005SSRv..120..165B} is a sensitive ($2\times 10^{-14}$ erg cm$^{-2}$ s$^{-1}à$ in $10^4$ seconds) broad-band (0.2-10\,keV) X-ray imager (23.6 arcmin field of view) with an angular resolution of 18 arcsec. Its flexible scheduling allowed us to observe Mrk\,421 regularly. The light curve was obtained from the online Swift-XRT products generation tool\footnote{http://www.swift.ac.uk/user\_objects/}, which uses \texttt{HEASOFT} software version 6.22.
The analysis is described in \cite{evans_2009MNRAS.397.1177E}.

\subsection{MAXI\label{par:maxi}}

MAXI is an experiment onboard the International Space Station, observing a large part of the sky every 96 minutes and most of the sky every day. Because of its higher sensitivity and sky coverage, we used the data from the Gas Slit Camera, consisting of 12 position-sensitive large-area counters sensitive from 2 to 20\,keV \citep{2009PASJ...61..999M}.

The light curve of Mrk\,421 was obtained by integrating the signal orbit by orbit (bins of 90 minutes) within a 9 square degrees region centred on the source. The background is estimated in square regions offsetted by $\pm 3$ degrees. The background-subtracted signal is normalised using the instrument response function \citep{2010PASJ...62L..55I} and made publicly available\footnote{http://maxi.riken.jp/star\_data/J1104+382/J1104+382.html}. 
The light curve of Mrk\,421 was binned on daily periods. Low-significance ($<2\sigma$) points, shown in red in Fig.~\ref{fig:LC}, were discarded  in the correlations studies.

\subsection{Optical and ultraviolet\label{par:optical}}

Mrk\,421 is monitored regularly in the V band in conjunction with Fermi observations of $\gamma$-ray-bright blazars \citep{2009arXiv0912.3621S}. The observations are carried out with the 1.54m Kuiper Telescope on Mountain Bigelow and the 2.3m Bok Telescope on Kitt Peak. For this study we used
publicly\footnote{http://james.as.arizona.edu/$\sim$psmith/Fermi/DATA/photdata.html} available data from Cycle 5 to Cycle 10 spanning from September 9, 2012, until March, 2018. 

Ultraviolet fluxes from Mrk\,421 are available from the Swift UltraViolet and Optical Telescope (UVOT) in three bands \citep[UVW1, UVM2 and UVW2,][]{roming_2005SSRv..120...95R}. The data were reduced with the HEASOFT package version 6.24 along with UVOT CALDB version 20170922. An aperture of 5 arcsec radius was used for the flux extraction for all bands. 
The background level was estimated in a circle of 15 arcsec radius located close to Mrk\,421 but excluding ghost images of the bright star 51 UMa, stray light, and UVOT supporting structures.

\subsection{Radio\label{par:radio}}

Radio observations of Mrk\,421 are performed regularly by the Owens Valley Radio Observatory (OVRO) 40 meter radio telescope as part of a blazar monitoring program \citep{Richards_2011ApJS..194...29R}. Observations were performed twice per week at 15 GHz (3 GHz bandwidth). The typical thermal noise reaches 4 mJy leading to a $\sim 3\%$ uncertainty. The data are available from the OVRO 40 meter telescope archive\footnote{http://www.astro.caltech.edu/ovroblazars/}. Of 329 data points, 8 with relatively large ($10-25\%$) uncertainties were kept as the fractional variability and correlation analysis take these uncertainties into consideration.

\section{Timing analysis\label{sec:timing}}

This section starts with an analysis of the variability amplitude of Mrk\,421 along the spectrum, providing some interesting information and followed by a study of the light curve auto-correlations and of the significant relations and responses found across the available wavebands.

\subsection{Variability\label{sec:fvar}}

The excess variance normalised by the flux, or the fractional variability of a light curve \citep[as proposed by][]{Vaughan_2003MNRAS.345.1271V}, can be estimated as $F_{var} = \sqrt{({S^2 - \langle\sigma^2_{err}\rangle})/{\langle{x}\rangle^2}}$
where S is the standard deviation of the light curve, $\langle\sigma^2_{err}\rangle$ is a mean squared error, and $\langle{x}^2\rangle$ is the average flux squared.
The uncertainties on $F_{var}$ are estimated following \cite{Poutanen_2008MNRAS.389.1427P} and \cite{Vaughan_2003MNRAS.345.1271V}. As data selection (no filtering, positive flux, or data with significance above $2\sigma$) has an effect on the fractional variability estimations (decreasing $\langle{x}\rangle^2$ and keeping $S$ unchanged), we 
calculated $F_{var}$ for all these cases separately.

\begin{figure}
  \centering
      \includegraphics[width=\columnwidth]{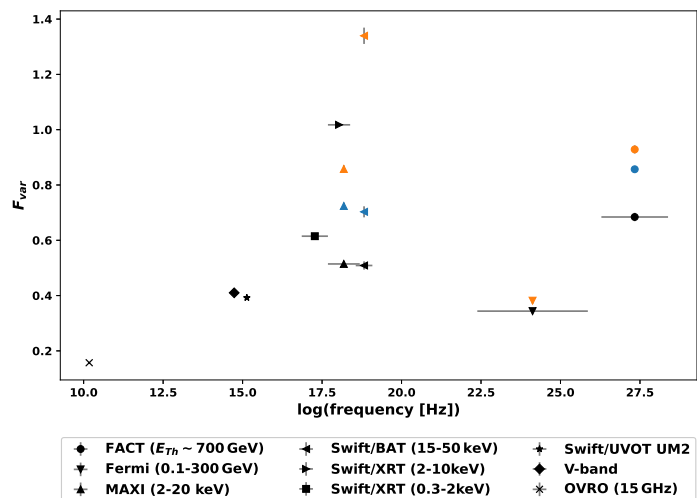}
  \caption{Fractional variability $F_{var}$ as a function of frequency. X-axis error bars indicate the energy band of the instrument. Y-axis error bars denote uncertainty on the $F_{var}$ value, and are  smaller than the marker for most instruments. For FACT, Fermi LAT, Swift/BAT, and MAXI, several markers are shown for different data-selection approaches: keeping all (orange), only $>2\sigma$ (black), or only positive (blue) data.}
  \label{fig:fvar_all}
\end{figure}

Figure \ref{fig:fvar_all} features the fractional variability of Mrk\,421 measured over 5.5 years across the spectrum. Similarly to previous studies \citep{2015A&A...576A.126A} the smallest fractional variability occurs in the radio and increases with energy up to the X-ray band. The fractional variability then drops to about 0.4 in the GeV (Fermi, 0.1-300\,GeV) and rises again towards the TeVs. We verified that different light-curve binning produced the same fractional variability, and therefore the results shown in Fig.~\ref{fig:fvar_all} represent a genuine property of Mrk\,421. During previous and shorter multi-wavelength campaigns on Mrk\,421 \citep[e.g.][]{aleksic_2015A&A...578A..22A}, the fractional variability was found to be about 0.55 in the TeV band, three times lower than shown in Fig.~\ref{fig:fvar_all}. This is most likely caused by the much ($10\times$) longer and unbiased light curves we used, which also include periods where the source is in high states. Similar $F_{var}$ values in X-ray and TeV bands were previously also reported for short-term observations, where the source was found in a very high activity state \citep{2020ApJS..248...29A}. The difference between the $F_{var}$ values for Swift/XRT (2-10\,keV) and MAXI (2-20\,keV), which have partly overlapping responses, can be explained by the fact that Swift/XRT probes variations on  shorter timescales.

The two maxima of the fractional variability of Mrk\,421 indicate that the high-energy portions of the two emission components are more variable than the low-energy ones. In addition, Fig. \ref{fig:xray_tev_flux} clearly shows  that the TeV and X-ray variations are linked and simultaneous, supporting the hypothesis that a single particle population is responsible for the emission in these two distinct energy bands.

It is interesting to compare this fractional variability with that found in Mrk\,501 \citep{ahnen_2017A&A...603A..31A}. In Mrk\,501, the fractional variability increases monotonically towards the TeV band, and reaches only about 50\% of that value in the X-rays. Mrk\,501 also features a lower TeV/X-rays correlation \citep{2015A&A...573A..50A} and a different high-energy hump spectral shape \citep{abdo_2011ApJ...727..129A}, suggesting an additional TeV emission component.
 
\subsection{Light-curve correlations\label{sec:crosscor}}

To study the correlation between irregularly sampled light curves we used the discrete correlation function \citep[DCF;][]{1988ApJ...333..646E}, which allows a time resolution independent of the data sampling. Uneven time spacing of data does not allow  the DCF time resolution to be decreased arbitrarily, unless long-term light curves are available, as in our case, providing enough correlation points within each time bin. We adopted one-day lag binning for the X-ray and TeV DCFs, and three-day lag binning for the DCFs involving radio, optical, and Fermi LAT light curves.

To cross check the DCF results we also calculated the Z-transformed DCF \citep[ZDCF;][]{Alexander_1997ASSL..218..163A}, which can more reliably detect correlations between sparsely time-separated data sets (each ZDCF bin has a different time width but the same number of points contributing). As our light curves are well sampled, the ZDCF did not improve the results.

We report uncertainties provided natively by the DCF. We did not use the \texttt{PSRESP} method  \citep[extension of the light curve;][]{2002MNRAS.332..231U} to estimate the DCF uncertainties as it occasionally amplifies correlations, especially when the power spectrum is not known.

We cross-correlated all light curves with each other and found two significant and interesting cases, reported in Sects. \ref{sect:tevx} and \ref{subsec:crosscor}.
When the FWHM of the DCF peak is larger or comparable to the data sampling period, the time lags can be estimated. To obtain the lag and its uncertainty,  we calculated the lag probability distribution from Monte Carlo simulations \citep[as in][]{Peterson_1998PASP..110..660P}. We generated $10^4$ subsets for each pair of light curves using flux randomisation (FR) and random subset selection (RSS) processes, and calculated the resulting DCFs to obtain a representative time lag distribution using a centroid threshold of 80\% of DCF maximum \citep{2004ApJ...613..682P}. The lag uncertainty corresponds to the standard deviation of the distribution of the lag values obtained for the random subsets.

We performed all auto- and cross-correlations filtering out data points of low significance (<$2 \sigma$ in general and removing all upper limits for Fermi/LAT), as often done in the literature \citep[e.g.][]{2011ApJ...738...25A}. By default the following figures and results are obtained with that selection. As removing periods when the source has a low significance (i.e. a low flux) can bias the results, we also performed the analyses keeping all data points (specifically mentioning these  in the text). Fortunately, the final results appear very similar in the two cases.

\subsection{Auto-correlations\label{sec:autocor}}

\begin{figure}[h!]
  \centering
      \includegraphics[width=\columnwidth]{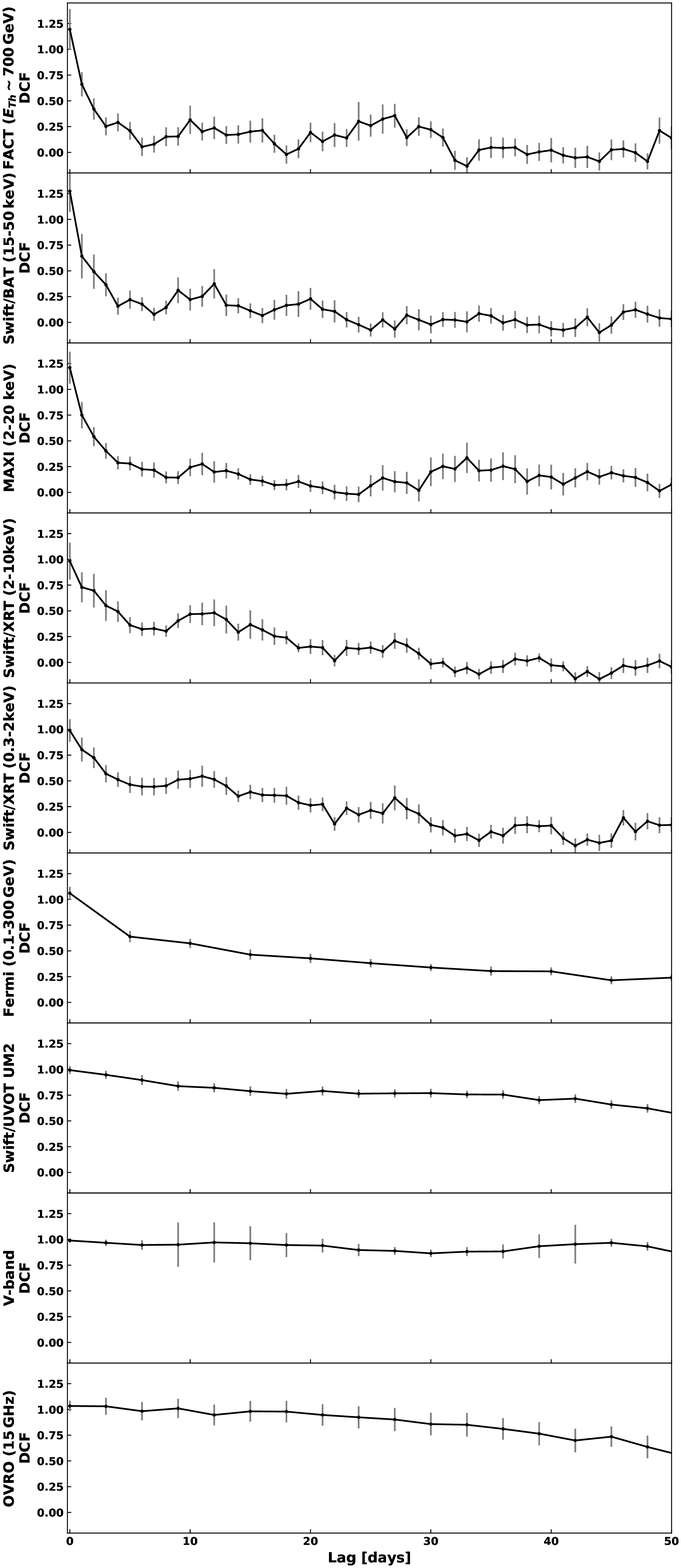}
  \caption{Light-curve auto-correlations. From top to bottom: FACT, Swift/BAT, MAXI, Swift/XRT (2-10\,keV), Swift/XRT (0.3-2\,keV), Fermi LAT, Swift/UVOT UVM2, V-band, and radio. Grey error bars denote 1$\sigma$ DCF uncertainties.}
  \label{fig:acf_all}
\end{figure}

The discrete auto-correlation function (DACF) is obtained by correlating light curves with themselves using the DCF. Following the same approach as for the cross-correlations above, we report the uncertainties as they are calculated by the DCF. The discrete auto-correlation functions of all available light curves of Mrk\,421 are shown in Fig.~\ref{fig:acf_all}, considering only >$2\sigma$ data points. A one-day DACF time binning was used for most data, excepting 5 days for GeV and 3 days for UV, optical, and radio data.

The  Fermi LAT fluxes have relatively high flux uncertainties, including correlated ones (the same sky model is used for each time bin). To obtain a proper normalisation of the auto-correlation, we chose a lag step of 5 days to have unity at zero lag, as described in \cite{1988ApJ...333..646E}.

The variability timescale observed in the TeV and X-ray bands is short (of the order of the binning timescale), as expected from models where the photons in the two bands are emitted by fast cooling electrons. Longer term responses ($\sim 10$ times the binning timescale) are observed in the GeV, optical, and radio, suggesting longer synchrotron cooling timescales at lower energies (see Sect. \ref{sec:discussion}).

\subsection{TeV--X-ray correlation\label{sect:tevx}}

\begin{figure}
  \centering
       \includegraphics[width=\columnwidth]{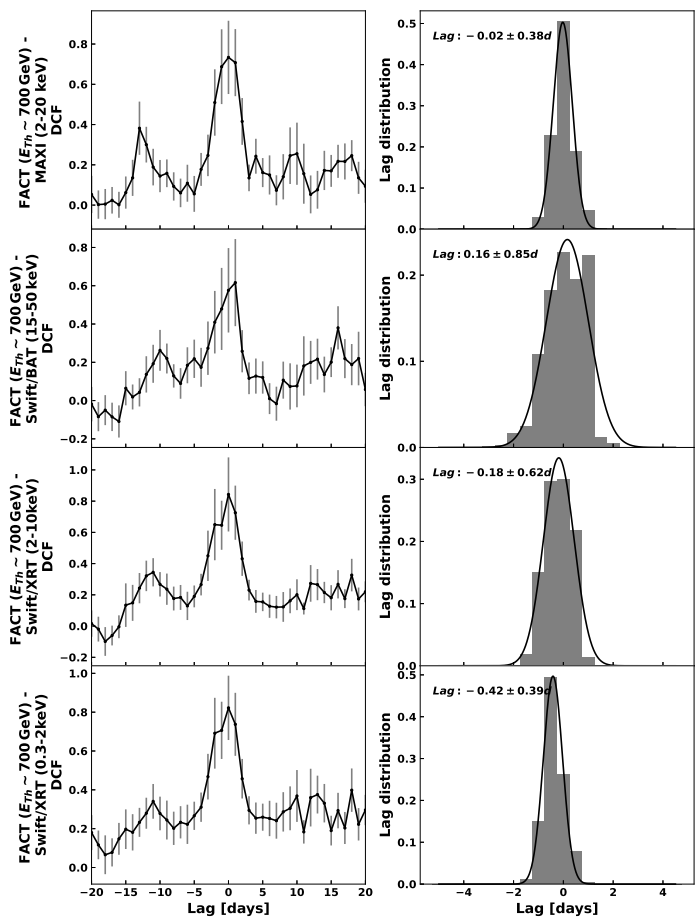}
       \caption{DCF cross-correlations of light curves (from top to bottom panel): FACT with Swift/BAT, MAXI, Swift/XRT (2-10\,keV), Swift/XRT (0.3-2\,keV). One-day binning was used. Left: DCF values as a function of lag. Grey error bars denote 1$\sigma$ uncertainties. Right: Lag distributions derived from FR/RSS simulations (more details are provided in Sect.~\ref{sec:crosscor}). A Gaussian fit (black lines) was used to derive the lag indicated on the plots.}
  \label{fig:dcf_all}
\end{figure}

\begin{figure}[!h]
  \centering
       \includegraphics[width=\columnwidth]{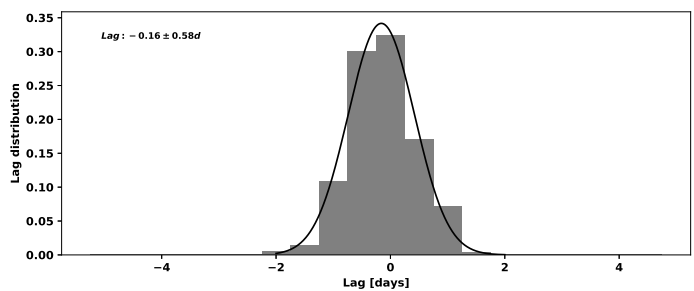}
       \caption{Combined lag distribution between TeV and X-rays (all bands) derived from FR/RSS DCF simulations (more details are provided in Sect.~\ref{sec:crosscor}). A Gaussian fit (black line) was applied to derive mean and uncertainty values.}
  \label{fig:combined}
\end{figure}

\begin{figure}
  \centering
       \includegraphics[width=\columnwidth]{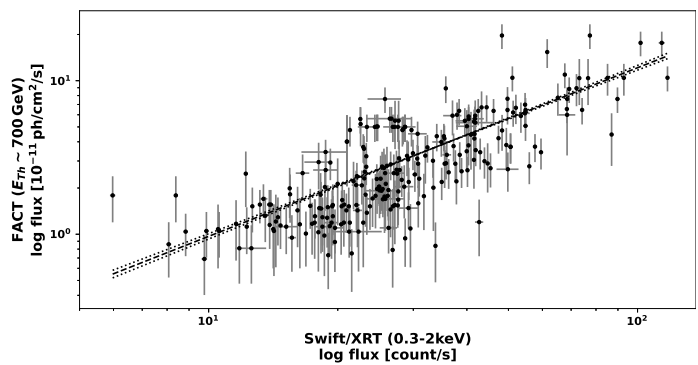}
       \caption{TeV (FACT) and X-ray (Swift/XRT 0.3-2\,keV) fluxes measured within 24 hours. The best-fit model is shown with a dashed line, and the  95\% confidence region is indicated with dotted lines.}
  \label{fig:xray_tev_flux}
\end{figure}

A strong correlation reaching 0.8 at zero lag is found between the TeV (FACT) and X-ray (Swift XRT, Swift BAT, MAXI) light curves (Fig. \ref{fig:dcf_all}). The $1\sigma$ upper limit on the lag is of the order of 0.6 days and summing the time-lag distributions corresponding to all X-ray light curves provides a combined lag of $(-0.16\pm0.58)$ days ($1\sigma$) (see Fig.~\ref{fig:combined}). Keeping all data points, whatever their significance, results in a combined lag of ($-0.17 \pm 0.48$) days.
The TeV/X-ray correlation was already reported using shorter and/or sparser data sets \citep{2005ApJ...630..130B,2015A&A...576A.126A,2016A&A...593A..91A} providing less constraining limits on the lag. The correlation between coincident (<24h time difference) X-rays (Swift/XRT 0.3-2\,keV) and TeV (FACT) fluxes can be seen in Fig.~\ref{fig:xray_tev_flux}. These lag limits are also compatible with the results of \cite{2008ApJ...677..906F} based on week-long light curves
and with the assumption that X-rays and TeVs are emitted by the same population of electrons. The dispersion of the data points in Fig.~\ref{fig:xray_tev_flux} around the main correlation (a slope of $1.098\pm0.033$ was obtained using the orthogonal distance regression method) could be explained by intra-day variability in these bands and/or by spectral variations between flares.

As the X-ray and $\gamma$-ray flares have a typical duration that is much shorter than the observing campaign we are reporting, the TeV/X-ray cross correlation can be checked flare-by-flare. To identify statistically significant changes of state automatically, and subsequently flares, we adopt the Bayesian Block algorithm \citep{2013ApJ...764..167S}. This allows us to divide the light curves into optimal blocks based on the variability statistical properties. The Bayesian Block algorithm identifies the points where the flux state is changing. The false-positive probability to define the state change was set to 1\% \citep{2013ApJ...764..167S}. As the Bayesian Block algorithm considers the uncertainties, the complete unfiltered light curves were used, as shown in Fig.~\ref{fig:LC}. We require each flare to last for at least two days and have an amplitude at least $2\sigma$ above the flux in the previous block. 

The Bayesian Block algorithm is sensitive to data sampling, flux variability, and flux uncertainties. As a result, the same change of the flux could be considered statistically significant or insignificant for two X-ray instruments operating simultaneously. We therefore considered all X-ray light curves and if a flare was detected in any of them, it was correlated to the list of TeV flares.

Out of 23 TeV flares, 22 were found to coincide with X-ray flares (Table ~\ref{tab:flares}). TeV and X-ray flares are of a similar duration. 
Removing all the low-significance ($\sigma<2$) measurements and running the Bayesian Block algorithm again decreased the number of coincident TeV/X-ray flares from 22 to 19 and removed the TeV-only flare.  The number of flares that disappeared (marked with asterisks in Table~\ref{tab:flares}) is in agreement ($\sim 20$\%) with the number of low-significance data points that were removed ($\sim 30$\%).

\begin{table}[ht]
\centering
\caption{List of TeV flares. Flares disappearing when low-significance measurements are removed are marked with an asterisk.}
\label{tab:flares}
\begin{tabular}[t]{lcp{4.3cm}}
Bands & Number & Time ranges, MJD\\
\hline
\hline
TeV only & 1 & 56320-56328*  \\ \hline
TeV, X-rays & 22 & 56370-56388, 56394-56397, 56696-56698*, 56981-57010, 57059-57069, 57087-57096, 57103-57116, 57123-57125, 57368-57385, 57385-57388, 57505-57511, 57533-57537, 57547-57550, 57754-57765, 57787-57789, 57802-57813, 57853-57864, 58105-58110, 58122-58134, 58134-58137, 58183-58199*, 58199-58201* \\ \hline
\end{tabular}
\end{table}

\subsection{Correlation with longer wavebands\label{subsec:crosscor}}

\begin{figure}
  \centering
      \includegraphics[width=\columnwidth]{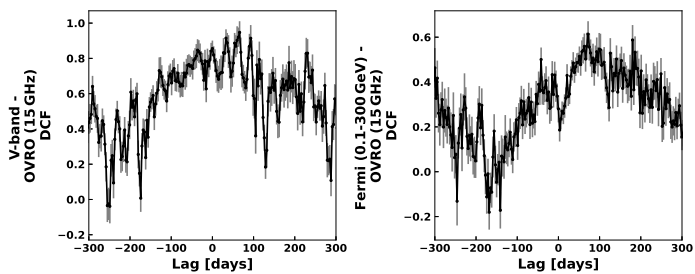}
  \caption{Left: Cross-correlation of the V-band and radio light curves. Right: Cross-correlation of the Fermi LAT and radio light curves. The time resolution is three days. Grey error bars denote 1$\sigma$ uncertainties.}
  \label{fig:cc_optical_radio_fermi_radio}
\end{figure}

The optical and ultraviolet light curves are highly correlated at zero lag despite their auto-correlation being wide. Both are also broadly correlated to the radio light curve (with a weaker coefficient at maximum) and are leading the radio variations by $\sim 30-90$ days at the maximum of the DCF (Fig. \ref{fig:cc_optical_radio_fermi_radio}, left).

The radio light curve is broadly correlated to the GeV light curve with a lag of $30-100$ days at the maximum of the DCF (Fig. \ref{fig:cc_optical_radio_fermi_radio}, right) and is not correlated to the TeV light curve, suggesting again separate parameters controlling the variability in the GeV and TeV bands. This delay, already observed and quantified to be $40 \pm 9$ days by \cite{max-moerbeck_2014MNRAS.445..428M} during a specific flare of Mrk\,421, was interpreted as the propagation of a shock through a conical jet. The same interpretation was proposed to explain the long-term light curves of 3C 273  \citep{turler_1999A&A...349...45T} and in particular the radio flares corresponding to overlapping stretched and delayed GeV flares \citep{esposito_2015A&A...576A.122E}.

\section{Discussion\label{sec:discussion}}

\subsection{Summary of results}

We performed a comprehensive study of the variability and of the correlations of multi-wavelength light curves of Mrk\,421 obtained by nine different instruments over 5.5 years from December 2012 to April 2018. The fractional variability of Mrk\,421 and the correlated TeV and X-ray emission are likely produced by a synchronous change of spectral shape of the low- (X-rays) and high-energy (TeV) components.

The excellent correlation at zero lag between the TeV and X-ray light curves of Mrk\,421 indicates that these two emissions are driven by the same population of high-energy particles. This variability could be driven by variations of the electron maximal energy, or by for example the magnetic field that would affect both electrons and protons \citep{Mannheim_1993A&A...269...67M}. The variability of Mrk\,421 is controlled, for instance, by the amplitude and the
shape of the electron energy distribution, which determine the fluxes observed on the left (radio, optical, GeV) or on the right (hard X-ray, TeV) sides of the two main components of the spectral energy distribution.

\subsection{Synchrotron self- or external Compton emission\label{sec:synchrotron}}

Several multi-wavelength campaigns on Mrk\,421 were conducted to determine the parameters of suitable SSC models from the observed spectral energy distributions \citep{1998MNRAS.293..239C, 2015A&A...576A.126A, abdo_2011ApJ...736..131A, aleksic_2015A&A...578A..22A}. Within the one-zone SSC model, the Doppler beaming factor $\delta$ and the magnetic field $B$ were estimated in the ranges $21-51$ and $(3-8)\times 10^{-2}$ G, respectively, with relatively high degeneracy because of the large number of parameters.

Some discrepancies were found between the observed spectra and those predicted by one-zone SSC models, especially for the Compton bump \citep{balokovic_2016ApJ...819..156B, 2015A&A...576A.126A}. Two-zone SSC models \citep{2015A&A...576A.126A} or more complex electron energy distributions \citep{zhu_2016MNRAS.463.4481Z} have been proposed to improve the situation.

Both the analysis of very fast X-ray variations during flares \citep{2015ApJ...811..143P} and that of low velocities in the high-resolution radio jet \citep{2013A&A...559A..75B} of Mrk\,421 have instead led to suggestions of a higher magnetic field $B\sim 0.5\,G$ ($\delta\sim 10$), contrasting with conclusions based on SED modelling. The cooling time for synchrotron-emitting electrons 
in the observer's frame of reference can be written as 
$t_{cool,\,e} \approx 15.86\times10^{11} \left((1+z)/\delta\right)^{1/2} \,(B/1{\rm G})^{-3/2} (\nu/1{\rm Hz})^{-1/2}\,{\rm seconds}$ \citep{2019ApJ...884..125Z}, where B, $\nu,$ and $z$ are respectively the magnetic field, the frequency of the observed synchrotron photons, and the source redshift.

In the high magnetic field regime ($B=0.5\,G$ and $\delta = 10$), the synchrotron cooling timescales become 7 minutes, 89 minutes, and 16.7 hours, for 50\,keV, 0.3\,keV, and in the optical, respectively, roughly in agreement with the fastest  variability timescale observed in Mrk 421 
\citep{2017ApJS..232....7F,2015ApJ...811..143P,1996Natur.383..319G,2011ApJ...738...25A}. 
In the low magnetic field regime ($B=0.1\,G$ and $\delta = 10$), the synchrotron cooling timescales in the observer frame become 1.2 hours and 16 hours 
for 50\,keV and 0.3\,keV, 
respectively, shorter than the flares that we are observing (typically 8 days, as listed in Table \ref{fig:flares_periods}). 
The average delay ($<0.6$ days) between the TeV and X-rays is in line with the SSC framework, where synchrotron and inverse Compton should be emitted simultaneously. 

\subsection{Lepto-hadronic and baryonic emission}

Lepto-hadronic and baryonic models may involve different baryonic or hadronic processes to explain the high-energy emission of blazars, while the low energies remain emitted by electron synchrotron.
The sychrotron--proton blazar model \citep{2001APh....15..121M} assumes that $\gamma$-ray photons are emitted by proton synchrotron. In Mrk\,421, $\gamma_{p,\, max} \approx 10^{10}$, $B\approx 50$ G and $\delta\approx 20$ are required to explain the observed TeV cutoff energy \citep{2013MNRAS.434.2684M, abdo_2011ApJ...736..131A}.  Synchrotron photons also interact with protons and induce pion cascades and muon synchrotron reaching energies $\gtrsim$ TeV . 

In the frame of the lepto-hadronic shock acceleration models, the correlation between the TeV and the X-ray flares indicates that electrons and protons are accelerated by the same process. However the proton acceleration time of $t_{acc}=20\,\xi\,\gamma\,m_p c/3\,eB \geq 10\,\xi\ {\rm days}$ \citep[where $\xi\geq 1$ is the mean free path in unit of Larmor radius,][]{2000ApJ...536..299K, 1996ApJ...463..555I}
is much longer than the delay observed between the TeV and X-ray light curves. Proton synchrotron therefore cannot be responsible for the TeV emission assuming the model parameters mentioned above (nor with smaller $B$ and higher $\gamma$). 

Muons, as secondaries of photo-protons interactions will also be generated on the proton acceleration timescales. Therefore, even if they could in principle be responsible for the TeV spectral energy distribution \citep{2017A&A...602A..25Z,abdo_2011ApJ...736..131A}, their production timescale is also much longer than the delay observed between the X-ray and TeV variations.

\cite{2016APh....80..115P} used a one-zone lepto-hadronic model to explain a 13-day flaring activity of Mrk\,421 (from X-rays to TeV, including intra-day variability). The variability timescale derived by these latter authors for the electron and proton luminosities also does not correspond to the timescales mentioned above for Fermi acceleration in shocks.
A different acceleration mechanism much faster than shock acceleration could be considered but this is out of the scope of the current paper.

\subsection{Hadronic emission}

Purely hadronic models invoke proton synchrotron for the low-energy component and  electron synchrotron from pion-induced cascades at higher energies \citep[$B\approx 20$ G \& $\delta\approx 16$,][]{2013MNRAS.434.2684M}. Such models were already disfavoured from the detailed study of the spectral energy distribution of Mrk\,421 \citep{2017A&A...602A..25Z}. 
The observed TeV variability timescale is compatible with leptonic emission but the synchrotron cooling timescale for X-ray-emitting protons in such conditions is extremely long with $t_{cool,\,p}\approx 1300\ {\rm years}$ \citep{2002MNRAS.332..215A} and even longer in the radio. Fast magneto-hydrodynamical changes could in principle explain the simultaneous X-ray and TeV correlation (as it affect all particles) but it is hard to imagine how the GeV-radio delay could be explained in this frame.

\subsection{Flare timing}

\begin{figure}
  \centering
      \includegraphics[width=\columnwidth]{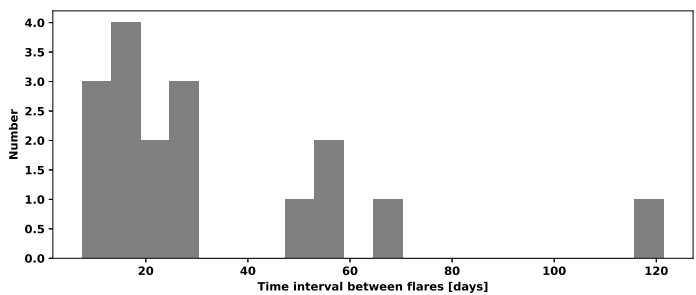}
  \caption{ Time-interval between the TeV flares listed in Table~\ref{tab:flares}. }
  \label{fig:flares_periods}
\end{figure}

\cite{Falomo_2002} estimated the mass of the black-hole in Mrk\,421 as $10^{8.5\pm 0.18}M_{\odot}$ from velocity dispersion measurements. 
The time interval $\Delta t$ between FACT-detected flares shows a broad distribution peaking between 7.5 and 30 days (see Fig.~\ref{fig:flares_periods}, the periods close to 60 days could be split in half, if weak or missed flares, not listed in Table~\ref{tab:flares}, are accounted for). The time interval in the source rest frame $\Delta t \cdot \delta / (1 + z)\sim 10^{3-4}\ R_G/c$ (where $\delta=10$) is comparable to that of the variations of the jet tilt \citep{10.1093/mnrasl/slx174} expected from Lense–Thirring precession \citep{1918PhyZ...19...33T} for an inclined accretion disk \citep{1975ApJ...195L..65B}. It is therefore possible that precession induces shocks travelling down the jet.

We did not find a relation between the TeV and X-ray light curves and those at GeV and radio. One scenario that could explain this observation is the following: the TeV/X-ray emission, being produced closer to the supermassive black hole, is dominated by variations in the electron maximum energy and the inner X-ray jet viewing angle, while the GeV and radio emission, being produced further downstream in the jet, are not affected by these parameters. Magnetic turbulence seems to result in a precession angle decreasing along the jet \citep[see Fig. 4 of ][]{10.1093/mnrasl/slx174}.

\section{Conclusions\label{sec:conclusions}}

This analysis of the multi-wavelength light curves of Mrk 421 obtained continuously with nine different instruments over 5.5 years provides two main observational results. The strongest variations of Mrk 421 occur in the hard X-rays and in the TeVs. X-ray and TeV flares are very well correlated (all the TeV flares were detected in the X-rays). The TeV and X-ray fluxes measured simultaneously (within 24 hours) are also correlated. Thanks to the long TeV light curve provided by FACT, the average lag between the TeV and X-ray variations could be estimated as $<0.6$  days ($1\sigma$). The GeV variations are strongly and widely correlated with optical and radio variability, with the latter lagging the GeV light curve by 30-100 days.

These constraints, together with the variability timescales observed in the various bands, were compared to the predictions of the three main classes of shock in jet models used to interpret the emission of blazars and of Mrk 421 in particular.
Purely hadronic models are ruled out by the very long variability timescales expected in particular in the X-ray band. Lepto-hadronic models predict proton acceleration timescales, which would imply delays of $\gtrsim$10 days between the X-ray and the TeV light curves, contrasting with the upper limit on the observed lag.

Electron synchrotron self- or external Compton models do match the observational constraints but require a delay to build up between the flares observed in the GeV and in the radio. Such a delay, which has been observed also in 3C 273 and  S5\,0716+714, could be related to evolution of the physical conditions when shocks move along the jet. As this delay has been observed combining the complete GeV light curve, it characterises a global property of the jet and not a idiosyncrasy of a specific shock.

\begin{acknowledgements}
Author contributions: V.S. and R.W. performed the combined analysis of multi-wavelength light curves and wrote the paper. M.B. performed the Fermi LAT analysis, D.D. performed the FACT analysis, V.S. performed the Swift/UVOT analysis and R.W. performed the Swift/BAT analysis. A.A.E., A.B., T.B., D.D., K.M. \& A.S. provided comments and discussed the results. All authors contributed to the construction and/or operations of the FACT telescope. The important contributions from ETH Zurich grants ETH-10.08-2 and ETH-27.12-1 as well as the funding by the Swiss SNF and the German BMBF (Verbundforschung Astro- 
und Astroteilchenphysik) and HAP (Helmoltz Alliance for Astroparticle Physics) to the FACT project are 
gratefully acknowledged. Part of this work is supported by Deutsche Forschungsgemeinschaft 
(DFG) within the Collaborative Research Centre SFB 876 "Providing Information by 
Resource-Constrained Analysis", project C3. We are thankful for the very valuable 
contributions from E. Lorenz, D. Renker and G. Viertel during the early phase of the FACT project. 
We thank the Instituto de Astrof\'{\i}sica de Canarias for allowing us to operate the telescope 
at the Observatorio del Roque de los Muchachos in La Palma, the Max-Planck-Institut f\"ur Physik 
for providing us with the mount of the former HEGRA CT3 telescope, and the MAGIC collaboration 
for their support. This research has made use of public data from the \textit{OVRO} 40-m telescope \citep{Richards_2011ApJS..194...29R}, the Bok Telescope on Kitt Peak and the 1.54 m Kuiper Telescope on Mt. Bigelow \citep{2009arXiv0912.3621S}, \textit{MAXI} \citep{2009PASJ...61..999M}, \textit{Fermi} LAT \citep{2009arXiv0912.3621S} and  \textit{Swift} \citep{2004NewAR..48..431G}. We would also like to express our gratitude to an anonymous referee.

\end{acknowledgements}

\bibliographystyle{aa}
\bibliography{mrk421_long_term}
\end{document}